\documentclass[lettersize,journal]{IEEEtran}
\usepackage{amsmath,amsfonts}
\usepackage{algorithmic}
\usepackage{algorithm}
\usepackage{array}
\usepackage[caption=false]{subfig}
\usepackage{textcomp}
\usepackage{stfloats}
\usepackage{url}
\usepackage{verbatim}
\usepackage{graphicx}
\usepackage{cite}
\usepackage{pifont}
\usepackage{booktabs}
\usepackage{amsmath}
\usepackage{amssymb}
\usepackage{amsthm}
\usepackage[utf8]{inputenc}
\usepackage{tcolorbox}
\usepackage{array}
\usepackage{multirow}
\usepackage{orcidlink}
\usepackage{booktabs}
\usepackage{tabularx}
\usepackage{makecell}
\usepackage{seqsplit}
\hypersetup{hidelinks}
\newcolumntype{C}[1]{>{\centering\arraybackslash}m{#1}} 
\newcolumntype{M}[1]{>{\centering\arraybackslash}m{#1}}

\makeatletter
\newcommand{\rmnum}[1]{\romannumeral #1}
\newcommand{\Rmnum}[1]{\expandafter\@slowromancap\romannumeral #1@}
\makeatother

\begin{document}

\title{Respond to Change with Constancy: Instruction-tuning with LLM for Non-I.I.D. Network Traffic Classification}

\author{Xinjie Lin$^{\orcidlink{0000-0003-0789-7570}}$, Gang Xiong,~\IEEEmembership{Member,~IEEE}, Gaopeng Gou, Wenqi Dong$^{\orcidlink{0009-0005-6177-0062}}$, Jing Yu, Zhen Li, Wei Xia

\thanks{Received 22 October 2024; revised 30 April 2025; accepted 20 May 2025. Date of publication 27 May 2025; date of current version 27 May 2025. This work is supported by The National Key Research and Development Program of China No. 2024YFF1401300. \textit{(Corresponding author: Jing Yu.)}}
\thanks{Xinjie Lin is with Zhongguancun Laboratory, Beijing 100094, China, and with the Institute of Information Engineering, Chinese Academy of Sciences and also with the School of Cyber Security, University of Chinese Academy of Sciences, Beijing 100085, China. (e-mail: linxj@mail.zgclab.edu.cn)}
\thanks{Gang Xiong, Gaopeng Gou, Wenqi Dong, Jing Yu, Zhen Li and Wei Xia are with the Institute of Information Engineering, Chinese Academy of Sciences, Beijing 100190, China, and also with the School of Cyber Security, University of Chinese Academy of Sciences, Beijing 100085, China. (e-mail: xionggang@iie.ac.cn; gougaopeng@iie.ac.cn; dongwenqi@iie.ac.cn; yujing02@iie.ac.cn; lizhen@iie.ac.cn; xiawei@iie.ac.cn)}
}

\markboth{Journal of \LaTeX\ Class Files,~Vol.~14, No.~8, August~2021}%
{Shell \MakeLowercase{\textit{et al.}}: A Sample Article Using IEEEtran.cls for IEEE Journals}

\maketitle

\begin{abstract}
Encrypted traffic classification is highly challenging in network security due to the need for extracting robust features from content-agnostic traffic data. Existing approaches face critical issues: (\rmnum{1}) Distribution drift, caused by reliance on the closed-world assumption, limits adaptability to real-world, shifting patterns; (\rmnum{2}) Dependence on labeled data restricts applicability where such data is scarce or unavailable. Large language models (LLMs) have demonstrated remarkable potential in offering generalizable solutions across a wide range of tasks, achieving notable success in various specialized fields. However, their effectiveness in traffic analysis remains constrained by challenges in adapting to the unique requirements of the traffic domain.
In this paper, we introduce a novel traffic representation model named Encrypted Traffic Out-of-Distribution Instruction Tuning with LLM (ETooL), which integrates LLMs with knowledge of traffic structures through a self-supervised instruction tuning paradigm. This framework establishes connections between textual information and traffic interactions. ETooL demonstrates more robust classification performance and superior generalization in both supervised and zero-shot traffic classification tasks. Notably, it achieves significant improvements in F1 scores: APP53 (I.I.D.) to 93.19\%(6.62\%$\uparrow$) and 92.11\%(4.19\%$\uparrow$), APP53 (O.O.D.) to 74.88\%(18.17\%$\uparrow$) and 72.13\%(15.15\%$\uparrow$), and ISCX-Botnet (O.O.D.) to 95.03\%(9.16\%$\uparrow$) and 81.95\%(12.08\%$\uparrow$).
Additionally, we construct NETD, a traffic dataset designed to support dynamic distributional shifts, and use it to validate ETooL's effectiveness under varying distributional conditions. Furthermore, we evaluate the efficiency gains achieved through ETooL’s instruction tuning approach.
\end{abstract}


\section{Introduction}
\IEEEPARstart{A}{s} an essential technology for cybersecurity and network management, traffic classification aims to identify categories of traffic from diverse applications or network services \cite{FuLSX21,AndersonM17,YeSWX22,LuoXLYXZHX22, YuanGZZXW24, RezaeiL19}, which has been widely used in scenarios such as security attack detection and quality of service assurance to help web content and service providers provide a more secure and high-quality web service experience for users.

In recent years, gradual full encryption of traffic has become a reality, explicit fingerprinting has been gradually failing. Different technical approaches have been proposed to address the needs of encrypted traffic analysis, including: (\textit{\rmnum{1}}) Statistical feature-based approaches \cite{PanchenkoLPEZHW16, TaylorSCM18} extract statistical features and combine them with classical machine learning algorithms to cope with traffic without plaintext; (\textit{\rmnum{2}}) Raw feature-based approaches \cite{LiuHXCL19, Sirinam2018} on the other hand selects raw traffic features and captures complicated patterns based on deep learning algorithms; and (\textit{\rmnum{3}}) Raw datagram-based approaches \cite{LinXG21, LotfollahiSZS20, WangZZYS17} utilize deep neural networks to learn implicit correlations between datagram bytes. 

Regrettably, the validity of most encrypted network traffic analysis methods is based on the assumption that training and testing traffic are independent and identically distributed (I.I.D.), following empirical error minimization learning from the training distribution. In fact, this assumption is fragile and unrealistic in practical scenarios in the field of cybersecurity. The interaction information and patterns of web applications change over time, which makes it difficult for existing methods to ensure good performance of the test data, the most intuitive manifestations of which include version updates of web applications, and behaviors in different temporal windows \cite{Liu0YFG21, GourdinMST17, Mobile2020}. Therefore, existing studies face the problem of probability distribution drift of traffic and category labels due to dynamic changes in network traffic, \textit{i.e.}, new feature distributions cannot be precisely mapped to the same labels under a well-trained classification model with the old distribution \cite{RimmerPJGJ18, JiangCLGXL23}.

\begin{figure}
	\centering
	\includegraphics[width = \linewidth]{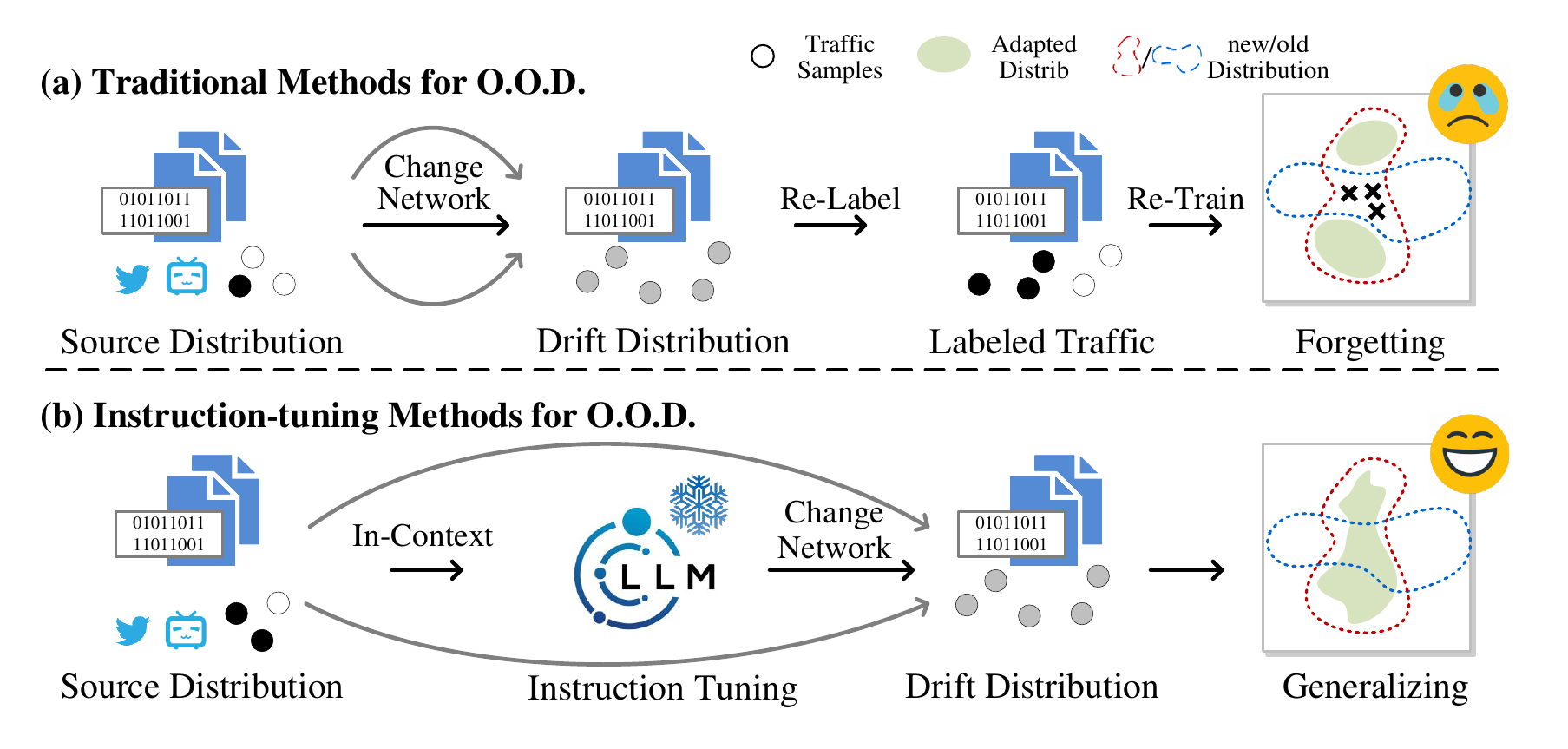}
	\caption{The Schematic Illustration of Different O.O.D. Identification Solutions.}
	\label{fig1}
\end{figure}

In response to the degraded performance of traffic classification models under Out-of-Distribution (O.O.D.) conditions, one of the most intuitive and commonly used techniques \cite{JuarezAADG14, Al-NaamiCMKLHT16, AttarianAH19} is to periodically retrain the model to adapt to changes in traffic, as shown in Fig. \ref{fig1}(a). However, updating the model involves collecting labeled samples and retraining the classifier, consuming a lot of time and labor. Moreover, the frequent updating of applications and the forgetting of old distributions make it difficult to balance the updating effort with performance degradation \cite{AttarianAH19, JiangCLGXL23}. Existing approaches are inappropriate in dealing with out-of-distribution traffic detection for two essential reasons:

(1) \textbf{Feature Instability}. Most used traffic features, single-packet features or single-flow features inherited from packet-level information, are weakly stable under distribution drift and lack the ability to represent traffic more robustly.

(2) \textbf{Insufficient Generalization}. Existing research frameworks are built to fit distributions under artificial experience or large-scale labeled data, while realistic networks, ranging from the complexity and variability of the application layer to the interactive changes of the transmission mechanism, launch an impact on the existing experimental assumptions.

Studies \cite{Al-NaamiCMKLHT16, TaylorSCM18, JiangLFCCXG22, WangYWGGYX24} have shown that both the temporal ordering of adjacent flows and packet-level bursts contribute significantly to traffic fingerprinting. Moreover, consistent traffic burst patterns within an application, even after updates, have proven to be robust features for building stable traffic representations. Additionally, pre-training methods are gaining traction in traffic analysis. For instance, ET-BERT \cite{LinXGLSY22} demonstrates strong generalization across multiple tasks, offering a viable traffic representation framework, though it does not fully resolve the generalization challenge.

Recently, large language modeling has been making breakthroughs in areas such as multi-modality. Under massive knowledge and unsupervised learning tasks, LLMs are able to acquire data extrapolation and scenario transfer capabilities, and emerge as emergent capabilities in some tasks \cite{WeiTBRZBYBZMCHVLDF22}. Such powerful generalization ability can be easily migrated to the target domain (Domain-LLM) \cite{TianGSZ024, YangZKXHA24} by fine-tuning the instruction form of small-scale labeled data. Stable traffic representation in conjunction with LLM motivated our idea.

To address the aforementioned challenges, we propose a novel traffic graph instruction tuning model for classifying encrypted traffic in out-of-distribution (O.O.D.) scenarios, called ETooL (\textbf{E}ncrypted \textbf{T}raffic \textbf{O}ut-\textbf{o}f-Distribution Instruction Tuning with \textbf{L}arge Language Model). ETooL focuses on designing flow interaction representations that allow LLMs to learn the underlying properties of network flows without requiring retraining for new distributions (Fig. \ref{fig1}(b)). First, we introduce flow graph representations, converting flow interaction properties into learnable graph structures. By using a contrastive learning approach, we align textual representations with flow graph structures, enabling the LLM to comprehend flow characteristics. During the instruction tuning phase, the model is guided through a BURST graph matching task as a self-supervised signal. This process helps the LLM understand the underlying transport structure (BURST), enhancing its ability to capture contextual associations in traffic. In the second tuning phase, we fine-tune the LLM using traffic-specific instructions, adapting it further to the traffic identification task.

In summary, the main contributions of this paper are summarized as follows:\par
\begin{itemize}
	\item[\Rmnum{1}]{We propose an instruction tuning model, called ETooL, for out-of-distribution encrypted traffic classification. The aim of this work is to align structural knowledge of the traffic domain with the generalization of LLMs, in order to enhance O.O.D. generalization for encrypted traffic.}
	\item[\Rmnum{2}]{We newly propose flow-specific self-supervised instruction tuning task, BURST Graph Matching, to improve the LLM's comprehension of flow interaction. Meanwhile, we introduce task-specific instruction tuning to enhance the adaptability to encrypted traffic classification.}
	\item[\Rmnum{3}]{We design and construct a dataset suitable for Non-I.I.D. traffic classification, named NETD. To the best of our knowledge, this is the first dynamically distributed traffic dataset that is dedicated to advancing data-supported research on O.O.D. encrypted traffic.}
	\item [\Rmnum{4}]{ETooL has great generalization ability and achieves a new state-of-the-art performance over 7 encrypted traffic classification datasets across independent and identically distributed and out-of-distribution scenarios, including  Encrypted Application Classification, Malicious Service Classification and Encrypted Traffic Classification with Distribution Flexible, and outperforms existing works remarkably by 6.62\%, 4.19\%, 18.17\%, 15.15\%, 9.16\%, 12.08\% and 2.88\%.}
\end{itemize}

\section{Preliminaries}
\label{sec:background}
\subsection{Problem Statement}
An adversary can use the encrypted traffic to perform a side-channel attacks to identify whether a victim has accessed a specific set of monitored applications. A defender, on the other hand, performs intrusion detection analysis with encrypted traffic to identify whether an attacker uses a malicious program to compromise a controlled network. We assume that the attacker or defender cannot exploit the plaintext payload of the packets and define an encrypted network flow as a bidirectional sequence of packets corresponding to a unique five-tuple {source IP, destination IP, source port, destination port, protocol}.

The goal of out-of-distribution encrypted traffic classification is to utilize traffic data from known distributions to learn transferable traffic knowledge, in order to achieve that the mapping relationship between test data and labels stays minimally changed when the distribution is changed, and thus to improve the accuracy of the traffic identification task under the new distribution. Specifically, we define the out-of-distribution encrypted traffic classification as follows:
Given the traffic samples in the data space $\mathcal{X}$ and label space $\mathcal{Y}$ as the initial data domain $D=\{(x,y) | x\in\mathcal{X},y\in\mathcal{Y}\} \sim P(x,y)$, the target data domain $D^{'}=\{(x,y) | x\in\mathcal{X},y\in\mathcal{Y}\} \sim P^{'}(x,y)$ is the newly distributed traffic data obtained by sampling with a different joint probability distribution from the initial data domain, then the learning objective ${f}_{\theta}: \mathcal{X} \rightarrow \mathcal{Y}$ is to maintain the accuracy of the label mapping in the event of a shift change in the marginal distribution of the traffic data:
\begin{equation}
	P'(Y|X) = P(Y|X), \text{if $P(X) \ne P'(X)$}
\end{equation}

\begin{table*}[!htbp]
	\centering
	\caption{The Comparison with the existing datasets of encrypted traffic classification.}
	\begin{tabular}{c|c|c|c|c|c}
	\toprule
        Dataset                 & Year & \#Flow  & Scenario   & O.O.D.    & Extensible \\ \midrule
        ISCX-VPN \cite{Draper-GilLMG16}               & 2016 & 2,329   & VPN/Service/Application             & \ding{56} & \ding{56} \\
        ISCX-Tor \cite{LashkariDMG17}                 & 2017 & 3,021   & Tor/Application            & \ding{56} & \ding{56} \\
        USTC-TFC \cite{WangZZYS17}                    & 2017 & 9,853   & Malware         & \ding{56} & \ding{56} \\
        Cross-Platform(iOS) \cite{EdeBCRDLCSP20}      & 2020 & 20,858  & Application             & \ding{56} & \ding{56} \\
        Cross-Platform(Android) \cite{EdeBCRDLCSP20}  & 2020 & 27,846  & Application           & \ding{56} & \ding{56} \\
        CSTNET-TLS 1.3 \cite{LinXGLSY22}              & 2022 & 46,372  & Service/Application         & \ding{56} & \ding{56} \\ \midrule
        ISCX-Botnet \cite{SamaniJSG14}                & 2014 & 96,857  & Malware         & \ding{52} & \ding{56} \\
        FDAN-APP53 \cite{JiangLFCCXG22}                & 2023 & 976,000 & Application            & \ding{52} & \ding{56} \\ \midrule
        NETD(Ours)                                    & 2024 & 3,000   & Service/Application        & \ding{52} & \ding{52} \\ \bottomrule
	\end{tabular}
	\label{table1}
\end{table*}

\subsection{Investigation on Existing Datasets}

Datasets are an invaluable component of contemporary traffic classification research, and they have been pivotal to the tremendous progress made in the field. Not only do they serve as a reliable source of training data, but they also provide a relatively fair means of measuring and comparing the performance of competing methods.  Due to the diversity of application scenarios and the development of network protocols, new traffic datasets are constantly being released to meet different research goals. However, most existing traffic datasets follow the assumption of independent and identical distribution of data, meaning that training and testing data should contain independent and identically distributed samples. In fact, we cannot decide the distribution of test data in real scenarios, and the assumption of independent identical distribution can never be strictly satisfied \cite{TorralbaE11}, which means that minimizing the empirical error of a model on the training data does not necessarily make it perform well on the test data.

Table \ref{table1} presents several representative datasets designed to meet various research objectives, covering scenarios such as VPN, malware, and mobile services. Most of these datasets were captured without fully considering the Non-I.I.D. nature of test scenarios, and only a few address distributional variations that are explicitly identified. For example, the ISCX-Botnet dataset introduces different data distributions by varying the type and volume of malicious and benign traffic between the training and testing sets. Similarly, FDAN-APP53 (APP53) accounts for both temporal and device factors to simulate distribution shifts caused by application version changes in real-world settings. The factors contributing to distribution shifts in real-world networks are complex, and it is labor- and time-intensive to comprehensively account for and capture them. For instance, in the case of temporal distribution shifts, constructing a dataset would require data collection over multiple time periods.

In this regard, we propose NETD, an out-of-distribution encrypted traffic dataset that supports distributional dynamics adjustment while being low-cost, efficient, and usable, as it supports the exploitation of publicly available datasets. It is worth noting that no previous dataset has supported adjusting the degree of traffic distribution bias, whereas the NETD dataset supports modelling varying degrees of O.O.D. traffic in a controlled manner. Details are given in Section \ref{sec:setting}.

\subsection{Motivation Analysis}

LLMs have demonstrated remarkable performance in tasks involving rich semantics and natural language understanding, highlighting their strong generalization capabilities. When it comes to encrypted network traffic, we apply LLM-based techniques based on the following considerations:

(1) \textbf{Feasibility of Applying LLMs to Encrypted Traffic Classification.} Although encrypted traffic lacks rich semantic information, the use of LLMs in encrypted traffic analysis is gaining momentum. Several studies \cite{LinXGLSY22,abs-2504-04222} have shown that LLM architectures (particularly leveraging pre-training) significantly enhance generalization in traffic classification tasks. By representing traffic features in a sequential format, LLMs can be naturally integrated to improve model generalization.

(2) \textbf{Effectiveness of Structured Traffic Graph Representations.} Structured graph representations of traffic have been proven effective in capturing stable interaction patterns between flows \cite{JiangLFCCXG22}. To further address the challenge of distributional shift in traffic data, we propose leveraging traffic graphs as an input representation for fine-tuning LLMs. Our experiments confirm that incorporating graph-based representations provides measurable gains in performance under distribution shift conditions.

(3) \textbf{Feasibility of LLMs Learning from Structured Data.} While LLMs are inherently designed for text, recent research \cite{JinLHJJH24} has demonstrated their ability to generalize across structured data formats. This inspired us to explore the integration of LLMs with graph-structured representations of network traffic, aiming to harness their transferability and generalization in non-textual domains.

(4) \textbf{Advantages of the LLM Architecture.} We evaluated LLM-based approaches using traffic sequence representations \cite{HeYC20,LinXGLSY22} alongside traditional AI techniques. Notably, most of these methods do not explicitly address the issue of distribution shift. Other existing approaches \cite{JiangCLGXL23} rely on access to unknown traffic in advance, which deviates from the strict requirements of true O.O.D. detection. In contrast, the LLM architecture exhibits stronger generalization capabilities, including zero-shot learning and representational transfer, which motivated our investigation.

\section{Related Work}

\label{sec:relatedwork}
In this section, we provide an overview of encrypted traffic classification methods that have been proposed, including statistical feature-based methods and deep learning-based methods, as well as pre-training and instruction tuning. Fingerprint construction represented by deep packet inspection (DPI) is no longer applicable and will not be discussed.\par

\subsection{Encrypted Traffic Classification}

\textit{(1) Statistical Methods:} To efficiently analyze complex traffic, most studies of encrypted traffic utilize statistical features of traffic independently of traffic encryption. CUMUL \cite{PanchenkoLPEZHW16} selects 104-dimensional statistical features by accuracy evaluation and then utilizes them as input to a support vector machine to identify website traffic. AppScanner \cite{TaylorSCM18} uses statistical features of packet size to train a random forest classifier. While ML-based methods combined with statistical features can analyze complex traffic, they rely on expert-designed statistical features, which makes it difficult to design generic statistical features to adapt to the large number of applications and websites that are constantly changing \cite{ShenLZXDG20}.

\textit{(2) Deep Learning Models:} Encrypted traffic classification using supervised deep learning in conjunction with raw features or raw datagram has become a popular approach to automatically extract distinguishing features rather than relying on manual design. FS-Net \cite{LiuHXCL19} uses recurrent neural networks (RNNs) to automatically extract representations from the original packet size sequence of encrypted traffic, while Deeppacket \cite{LotfollahiSZS20} and TSCRNN \cite{LinXG21} are representing the original payload. Traffic Interaction Graph \cite{ShenZZXD21} models flow interactions as graphs and learns flow associations based on graph representations, providing better traffic identification ability. However, such methods rely on a large amount of supervised data to capture the effective features and thus learn biased representations in a small range of data.

\subsection{Pre-training and Instruction Tuning}

Pre-training techniques learn unbiased data representations from large amounts of unlabeled data through self-supervised learning, which not only significantly reduces the appetite for labeled training data, but also further improves performance in downstream tasks. In encrypted traffic classification, pre-training models are applied as emerging architectures to improve the generalization of traffic classification. PERT \cite{HeYC20} applies the pre-training model to migrate ALBERT to encrypted traffic classification and achieves performance improvement in VPN scenarios. ET-BERT \cite{LinXGLSY22} proposes pre-training tasks that are more suitable for traffic datagram representation and achi\-eves performance improvement in traffic classification under multiple tasks, which demonstrates the powerful generalization of the pre-training model for encrypted traffic classification. In addition, MT-FlowFormer \cite{ZhaoDYMXW22} and Flow-MAE \cite{HangLWX23} utilize the pre-training model to capture flow correlations from a visual perspective and improve flow identification performance. Pre-training techniques demonstrate power in traditional traffic classification, but these studies fall short of the desired goal by not considering attempts to solve the problem of classifying Non-I.I.D. encrypted traffic.

Prompting \cite{PetroniRRLBWM19}, a technique that uses task-specific prompts to guide pre-trained models, reducing the need for fine-tuning or large amounts of labeled data. Recently, prompt learning has demonstrated its effectiveness in generalized transfer in natural language processing, computer vision, and network service optimization tasks \cite{ZhouYLL22,WuWQ0JC024}. The instruction tuning paradigm, integrating the pre-train and fine-tuning framework with prompt learning, enhances generalization capabilities in transfer learning by enabling effective task adaptation with few or even zero samples \cite{Tang00SSCY024, WangKMLSKH23}. In the context of encrypted traffic classification, this area of research remains largely unexplored. We propose a generic traffic representation based on domain-specific traffic knowledge, taking into account the unique characteristics of traffic modalities. Additionally, we design two instruction-tuning tasks to ensure the generalized transferability of traffic representations.

\section{Overview of ETooL}
\label{sec:methodology}
In this section, we present the design of ETooL. Typically, encrypted traffic analysis focuses on extracting multi-dimensional features from a single data flow and examines the flow pattern under the assumption of I.I.D. data. However, this assumption is often fragile and unrealistic in real-world scenarios. Single-flow patterns are more susceptible to performance degradation due to distributional bias compared to multi-flow interaction patterns.

\begin{figure*}
	\centering
	\includegraphics[width = \linewidth]{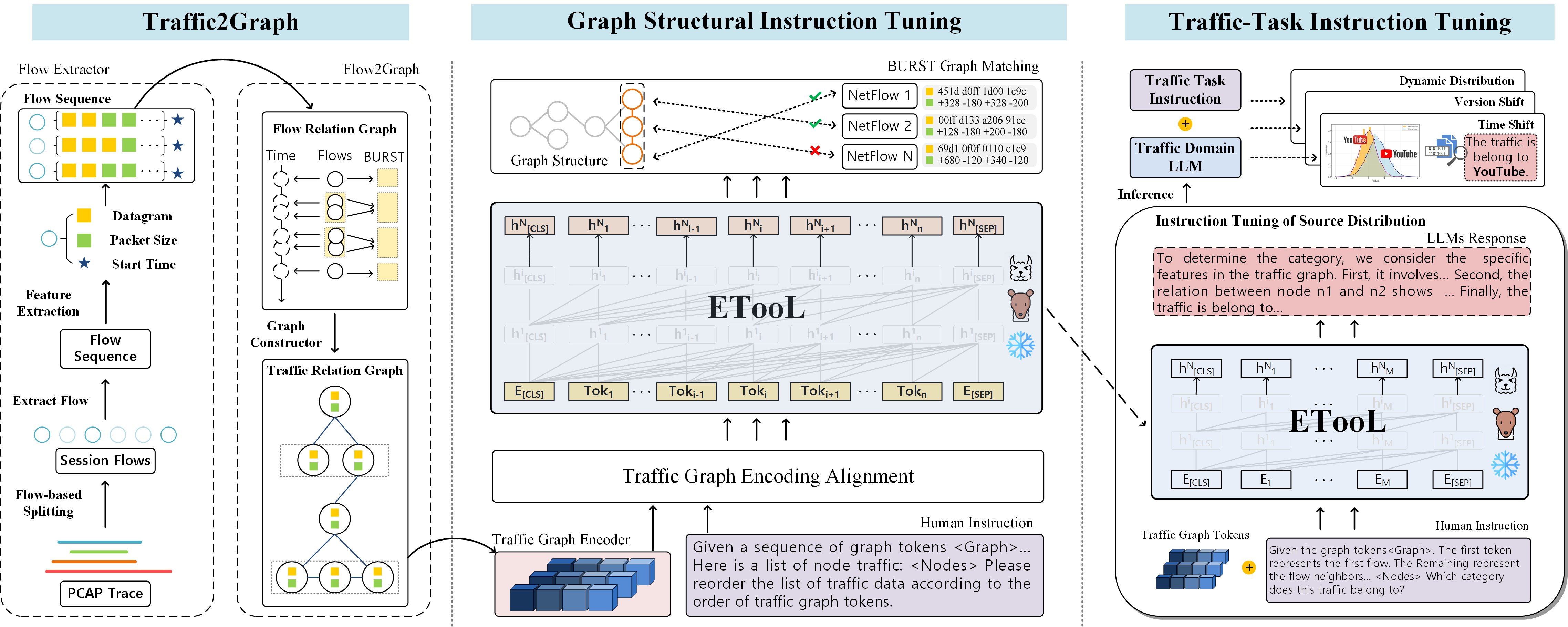}
	\caption{Overview of ETooL Framework.}
	\label{fig2}
\end{figure*}

To this end, our goal is to learn generic interaction correlation patterns for encrypted traffic and achieve better netflow classification in scenarios with different distributional variations. Thus, in this paper, we propose ETooL, an out-of-distribution encrypted traffic classification framework based on a generalized pre-trained large model, to tackle the out-of-distribution recognition problem in encrypted traffic in the network domain. As shown in Fig. \ref{fig2}, the ETooL framework consists of a two-stage fine-tuning and contains three core components, \emph{i.e.}, traffic interaction graph structure representation, graph structure instruction tuning, and traffic task instruction tuning.

\textbf{Traffic2Graph.} Drawing inspiration from FRG features \cite{JiangLFCCXG22} that cope with ambiguous flows and concept drifts, we propose to utilize flow interactions incorporating multi-granularity features as a generic pattern for constructing flow representations. Network Traffic Relation Graph (TRG) is constructed based on the correlation topology between different flows, and the graph contains flow features at different granularities. Specifically, the TRG consists of multiple network flows at adjacent timing, where each node represents a network flow, and each node contains a Raw Datagram (RD) sequence and a Packet Length (PL) sequence. The relationship between nodes represented by different network flows, \emph{i.e.}, correlation edges, consists of \textit{adjacency edges} and \textit{burst edges}, which represent the interaction between different flows.

\textbf{Graph Structural Instruction Tuning.} This phase proposes a well-designed flow-graph alignment module and a flow-graph structure instruction tuning paradigm for helping LLMs capture and learn flow associations, thus alleviating problems such as the difficulty of existing LLMs in understanding flow feature information and flow-graph structure. In particular, the flow-graph alignment module aims to align the flow features and topological relationship graphs in the encoding space, based on which natural language instruction data containing flow feature information is designed for self-supervised tuning, leading to better understanding of flow graph structure knowledge.

\textbf{Traffic-Task Instruction Tuning.} In order to help the large language model adapt to the out-of-distribution traffic identification task, this phase proposes the traffic task instruction tuning module to design the instruction data for traffic classification based on the knowledge of the traffic domain structure obtained by the large language model.

We realize the learning of network interactions and contextual associations by constructing correlation graphs to represent the interaction characteristics of multi-flow, and allowing the pre-trained models to recognize the flow correlation patterns through self-supervised learning. Meanwhile, with the inference and understanding ability of the pre-trained model, we realize the generalized out-of-distribution flow identification under the traffic task instruction tuning.

\section{Traffic2Graph}
\label{sec:t2g}

In real networks, multiple network flows are often established within a short timeframe to enhance the application's response rate and improve user experience. This results in several communication flows being created and transmitting messages simultaneously, which can be observed through passive network traffic capture.

This phenomenon is referred to as BURST, a key concept widely used in recent years for traffic feature mining and representation. Specifically, \textit{packet-level BURST} is defined as a sequence of consecutive packets whose arrival intervals are within a small time threshold. Several studies have demonstrated that this traffic structure is effective for analyzing encrypted network traffic. Similarly, \textit{flow-level BURST} refers to network flows established within a short time window, which helps explore the correlation properties between flows. While packet-level BURST captures the traffic characteristics of multiple packets serving the same resource request or response, flow-level BURST reflects the collaboration between multiple flows serving the same network function.

To further capture the interactions between flows in raw traffic, we propose the Traffic2Graph module to construct a discriminative and generic traffic characterisation with the variability that exists in the flow-level BURST features of different web applications. The module consists of two processes: (1) Flow Extractor extracts datagrams and packet sizes from each input network flow to merge flow feature information of different dimensions. (2) Flow2Graph further constructs the flows with fused features into a graph structure according to timing and connectivity relationships to better represent the correlation information among different flows in BURST.

Web application developers often implement similar functionalities, but their differing interpretations of business models and development practices can result in distinct network flow collaboration patterns at the network traffic level. This section examines the interaction between network flows, focusing on two key relationships: adjacency and bursting. Adjacency describes the connectivity between adjacent network flows and, in this context, is extended to represent the through-connectivity between neighboring BURST structures. The bursting relationship, on the other hand, refers to the connectivity between flows within the same flow-level BURST structure. These multi-dimensional correlations capture the temporal and sequential relationships between network flows and the collaborative construction of BURST structures, which work together to enable network service functionality.

\subsection{Flow Extractor}
The multi-dimensional fusion of traffic information helps to adapt to the needs of more traffic identification scenario tasks, and we focus on datagrams and packet sizes that are widely used and have effects: datagram and packet size. In particular, to adapt to flow data in instruction fine-tuning, we expand the token representation of LLM according to CETP \cite{LinHGYGLGX24}.

(1) Sequence of Datagrams. We extract 128 bytes from the datagrams and construct traffic representation units that contain more information to enable the pre-training phase to obtain richer contextual information. Therefore, the hexadecimal bit sequence in the original traffic datagram is double-byte split and encoded as a sequence of byte pairs, where the representation space of each unit ranges from 0 to 65535, \emph{e.g.}, the \{ee08bf56...\} would be represented as \{ee08, bf56, ...\}.

(2) Directed Packet Size Sequence. The construction of the packet size sequence follows the conventional way of flow statistical characterisation, where the packet size is extracted while preserving the communication direction information of the encrypted flows, where + indicates that the packet is sent from the client to the server, and - indicates that the packet is returned from the server to the client. For example, a sequence of directed packet sizes for a bi-directional flow can be represented as \{+128, -74, -1020, +378...\}.

\subsection{Flow2Graph}
Given all network flows S generated by a client using a certain web application during a certain period of time, and a graph structure is constructed for these flows. Taking advantage of the property that graph data structures can express rich node information and relationships between disjoint nodes, the traffic relation graph TRG$(G=(V, E))$ is used to express adjacency and bursting relationships between different network flows. The specific construction process of the traffic relation graph is shown in Algorithm \ref{alg:alg1}.

In accordance with the adjacency and bursting edges between flows in the temporal relationship, and the feature information representation in each flow, we construct the nodes and edges in the traffic relation graph as follows:

(1) Nodes $V$ in TRG. Each network flow constitutes a node in the TRG, where each node consists of flow features including datagram sequences, packet size sequences, packet message type sequences, packet time interval sequences, and so on. Since the value of the start timestamp of a network flow is dynamically variable, we are mainly concerned with the directed packet size sequence and datagram sequence of each network flow.

(2) Edges $E$ in TRG. The TRG contains two types of association edges, the first of which is adjacency edge for connecting flows within different BURST structures. By capturing the bursting relation, the flows in the set $S$ can be divided into different BURST structures, after which the neighbour relation is formed by connecting the last flow of the previous flow level BURST to the first and last flows of the next BURST. And the second one is the burst edge, which is applied to connect concurrent network flows within the same BURST structure. The flow level BURST is divided according to whether the start timestamp of the network flow is within a small temporal neighbourhood $\gamma$.

\begin{algorithm}[H]
\caption{Construction of Traffic Relation Graphs.}\label{alg:alg1}
\begin{algorithmic}[1]
\REQUIRE $S=\{f_1,f_2,...,f_n\}$: Network traffic data collected over time; $\gamma$: Interval threshold for determining the flow level BURST;
\ENSURE $G=(V,E)$: The traffic relation graph $G$ containing nodes $V$ and edges $E$;
\STATE $V=\{\}$, $E=\{\}$, BURST=$[ \ ]$, BURST$_{last}=[ \ ]$
\STATE Sort the network flows in $S$ by starting timestamps
\STATE Each flow is added as a node to $V$
\FOR {each $f \in S$}
    \IF{BURST $\neq NULL$}
        \IF{$|f_{start\_time} - \text{BURST}^{last}_{start\_time}| \leq \gamma$}
            \STATE current flow $f$ is added to BURST
        \ELSE
            \FOR {$i \in range(\text{BURST}_{size})$}
                \STATE $E.insert(\text{BURST}[i],\text{BURST}[i+1])$
            \ENDFOR
            \IF{$\text{BURST}_{last} \neq NULL$}
                \STATE $E.insert(\text{BURST}_{last}[-1],\text{BURST}[0])$
                \STATE $E.insert(\text{BURST}_{last}[-1],\text{BURST}[-1])$
            \ENDIF
            \STATE $\text{BURST}_{last} = \text{BURST}$
            \STATE $\text{BURST} = [ \ ]$
        \ENDIF
    \ENDIF 
\ENDFOR
\STATE \textbf{return}  $G=(V,E)$
\end{algorithmic}
\label{alg1}
\end{algorithm}

\section{Graph Structural Instruction Tuning}
\label{sec:gsit}
To enhance the understanding of flow graph structural information with LLMs, ETooL aligns the flow graph structural encoding with the natural language space. This alignment is intended to enable the language model to leverage the inherent language comprehension capabilities for effective understanding of flow features and through-connection relationships. Towards this goal, we design a flow graph encoding alignment module that aims to preserve the flow graph structural context information during instruction tuning of the large language model, thus effectively correlating the flow understanding with the topological structure relationships in the graph.

\subsection{Traffic Graph Encoding Alignment}
Inspired by cross-modal alignment studies such as CLIP, we integrate traffic features into the graph structure encoding process in the form of contrastive learning to align and fuse the traffic graph structure and traffic information representation.

Specifically, a graph neural network encoder with pre-training parameters is integrated into the ETooL framework and enabled to correspond to the graph representation and the flow representation encoding through the contrastive learning approach. Assuming that the flow graph is represented as $G(V,E,A,X)$, the flow feature information of the nth node corresponding to the flow is represented as $C = \{c_i \in \mathbb{R}^{l_i \times d}, 1 \leq i \leq N\}$, where $l_i$ denotes the length of the input of the $i_{th}$ node and $N$ denotes the number of nodes.

The encoded flow graph structure and flow feature representations are obtained by any graph representation encoder $f_g$ (\emph{e.g.}, Graph Transformer) and flow representation encoder $f_n$ (\emph{e.g.}, ET-BERT) as follows:
\begin{equation}
	H = f_g(G), N = f_n(C)
\end{equation}
where $V$,$E$,$A$,$X$ as inputs to the flow graph denote the node encoding, associated edges, graph adjacency matrix, and node features, respectively. $H$ denotes the structure-level graph encoding representation generated by the graph neural network encoding, and $N$ is the encoded representation of the flow features associated with the nodes.

The traffic-graph alignment process for different dimensions through comparative learning is conducted as follows:
\begin{equation}
	\Im_i = (g_i^1(norm(H)) \cdot g_i^2(norm(N))^{\top}) \cdot \exp(\tau)
\end{equation}
\begin{equation}
	L = \sum_i{\frac{1}{2}\lambda_i(\text{CE}(\Im_i,y) + \text{CE}(\Im_i^{\top},y))}
\end{equation}
where $y=(0,1,...,-1)^\top$ as the contrastive learning target denotes the alignment label, $\Im_i$ denotes the similarity measure during contrastive learning, $g_i$ denotes the encoder of different information, $\tau$ denotes the temperature coefficient, $\lambda$ denotes the weight coefficient of different difficulty sample pairs, and CE denotes the cross-entropy loss function.

\subsection{BURST Graph Matching}

The encoding alignment enables the inclusion of flow features in the instruction cues to be understood in association with the flow graph structure. In order to further align the linguistic comprehension of the large language model with the graph learning task, we utilise a pre-trained instruction tuning paradigm to enhance the adaptability of the large language model for specific traffic learning tasks, enabling the large language model to generate more accurate and contextually appropriate results for the traffic graph structure data.

Despite the strong generalised contextual understanding of the large language model, it is lacking in the understanding of network communication behaviours as well as traffic features. Meanwhile, the construction of traffic understanding capability is independent of a specific traffic identification task, and we use self-supervised instruction tuning to inject the knowledge of traffic graph structure into the large language model so as to effectively understand the contextual information in the traffic. Specifically, we design the interactive structure-aware flow graph matching task in the self-supervised instruction tuning approach, which allows the use of unlabelled encrypted traffic data to generate the representations of the flow graph structure as part of the instructions for the tuning of the large language model, where the flow graph structure will be used as a self-supervised signalling unit to instruct the large language model to distinguish between the different flow graph nodes using both the natural language and the flow sequence.

\textbf{Instruction Design.} The traffic graph matching task is guided by three core components: (\textit{\rmnum{1}}) traffic graph, (\textit{\rmnum{2}}) problem instruction, (\textit{\rmnum{3}}) ETooL response. Each node within the traffic graph is designated as a central node, and an h-hop random neighbor sampling strategy is employed to extract the subgraph structure from the input traffic graph. The input provided to the large language model is a combination of natural language descriptions and traffic-specific features. For this task, the instructions consist of a $<$graph$>$ indicator unit, disrupted BURST traffic features, and a textual problem description. The objective is to align each flow, represented by a traffic graph node, with its corresponding traffic feature. Achieving this alignment necessitates reordering the disordered BURST traffic feature representations by understanding the topological relationships between graph nodes. Through this process, the model correlates the structural representation of the traffic graph with its associated traffic features, thereby improving its capacity to infer and comprehend network traffic behavior.

\textbf{Tuning Strategy.} Through the lightweight alignment projection, we keep the parameters of both the LLM and the graph neural network encoder frozen during tuning, optimizing only the parameters within the projection layer. Specifically, we freeze all components of the LLM backbone, including attention blocks, token embeddings, and layer normalization layers, as well as every layer within the pre-trained flow-graph encoder. After fine-tuning, the projection layer effectively maps encoded flow-graph representations to corresponding node representations, enabling the LLM to align these node representations with various node-level feature semantics. By using a projector (\emph{e.g.,} a linear mapping module), the model establishes a correspondence between graph node representations and flow feature instruction representations. The indicator unit $<$graph$>$ embedded in the natural language instructions is replaced with the aligned flow graph node representations, formatted as $\{<$graph\_begin$>$, $<$graph\_token$>_1$, $...$, $<$graph\_token$>_n$, $<$graph\_end$>\}$. This incorporation of flow graph structure into the instructions allows the large language model to process them. Since the flow graph matching process is self-supervised, it efficiently leverages large amounts of unlabeled flow graph data from various traffic scenarios, improving the generalization capabilities of the learned projectors.

\section{Traffic-Task Instruction Tuning}
\label{sec:tit}
After completing self-supervised instruction tuning, we develop task-specific instruction tuning methods tailored to encrypted traffic classification. This process involves customizing the inference behavior of the LLM to meet the specific constraints and requirements of the classification task. By fine-tuning the LLM with task-specific instructions, the model is guided to generate responses that are more appropriate for traffic learning. The traffic task instruction tuning further enhances the model's adaptability in handling encrypted traffic identification tasks across varying distributions.

In the traffic task instruction tuning, the instruction template consists of three parts. The traffic graph information includes multiple traffic samples collected over time, supporting the construction of sub-graphs. For the traffic classification task $\phi$, we model the traffic representation using the training pair $(X, y)$. During this phase we continue to keep the full LLM backbone and the flow-graph encoder frozen, updating only the structure-aware projector inherited from BURST Graph Matching together with a lightweight task-specific classification head. Then ETooL leverages the parameters of the structure-aware projector, trained in the first phase as the initial state $\theta$, and fine-tunes it to predict the traffic label $y$.
\begin{equation}
	y = ETooL(X | (\theta;\phi))
\end{equation}

After completing the dual-stage instruction tuning, which includes freezing specific model parameters, the large language model's ability to understand and infer the structure of traffic graphs is significantly improved. This approach enables the model to efficiently handle a wide range of tasks related to traffic graph analysis.

\begin{table*}[htbp]
    \centering
    \caption{The Statistical Information of Datasets. The actual number of datasets used for classification after processing. The categories and sample size of NETD are determined by the actual dataset used.}
    \label{tab:datasets}
    \begin{tabular}{c|c|c|c|c|c}
    \toprule
    Tasks& Description of Task& Dataset& \#Flow& \#Label & Collection\\
    \midrule
    EAC-T & {Encrypted Application Classification with the Same Time Distribution} & APP53-TIME\cite{JiangLFCCXG22} & 93,479 & 28& Public\\
     \midrule
    EAC-V & {Encrypted Application Classification for the Same Application Version} & APP53-VERSION\cite{JiangLFCCXG22} & 104,697 & 25& Public\\
    \midrule
    EAC $\Rightarrow$ T & Encrypted Application Classification with Time Shift & APP53 $\Rightarrow$ TIME\cite{JiangLFCCXG22} & 93,479 & 28& Public\\
    \midrule
    EAC $\Rightarrow$ V & Encrypted Application Classification with Version Shift & APP53 $\Rightarrow$ VERSION\cite{JiangLFCCXG22} & 104,697 & 25& Public\\
    \midrule
    MSC $\Rightarrow$ T & Malicious Service Classification with Type Shift & ISCX-Botnet\cite{SamaniJSG14} & 8,000(5,578) & 2(8) & Public\\
    \midrule
    ETC $\Rightarrow$ F & Encrypted Traffic Classification with Distribution Flexible & NETD & - & - & Private\\
    \bottomrule
    \end{tabular}
\end{table*}

\section{Evaluation}
\label{sec:evaluation}
In this section, we perform six different encrypted traffic classification scenario tasks (Section \ref{sec:setting}) to demonstrate that ETooL has better generalisation and effectiveness in the out-of-distribution traffic identification task, as well as to show that the ETooL model can still be adapted to the old distributed encrypted traffic identification task. We then compare our model with 6 approaches (Section \ref{sec:exp1}) and perform an ablation analysis of the key components of the model (Section \ref{sec:exp2}). We further provide an analysis of the effectiveness of ETooL for traffic identification in a dynamic distribution offset scenario (Section \ref{sec:exp3}), an evaluation of the efficiency of the model (Section \ref{sec:exp4}), as well as the analysis of hyper-parameter selection (Section \ref{sec:exp5}).

\subsection{Experimental Settings}
\label{sec:setting}

\subsubsection{Datasets Descriptions}
\label{sec:dataset}
To evaluate the effectiveness and generalization of ETooL, we conduct experiments across five encrypted traffic classification tasks on four public datasets and one one newly proposed dataset. The tasks and the corresponding datasets are shown in Table \ref{tab:datasets}.

We conduct I.I.D. and O.O.D. experiments in the publicly available dataset APP53 \cite{JiangLFCCXG22}, which contains the 53 web apps with the largest user sizes selected from the Google Apps Marketplace, and collect data from these apps across time and versions on different devices by volunteers. However, since this dataset only exposes the encrypted traffic dataset of apps collected on Xiaomi 5Plus devices at different times and with different versions, the following task will be set up around that: extracting only apps from APP53 that have undergone version changes and time changes, and setting them up as Independent Identically Distributed and Non-Independently Identically Distributed experiments.

\begin{figure}[t]
    \centering
    \subfloat[ISCX-VPN]{%
        \includegraphics[width=\columnwidth]{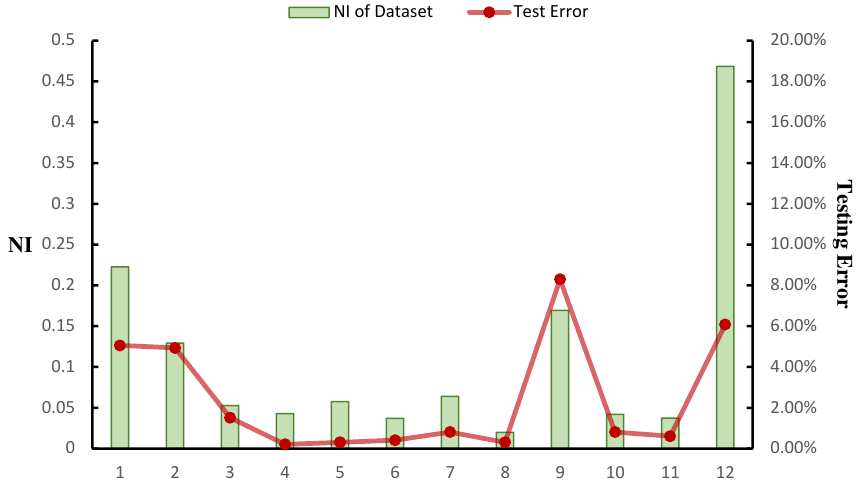}%
    }

    \subfloat[NETD-1]{%
        \includegraphics[width=\columnwidth]{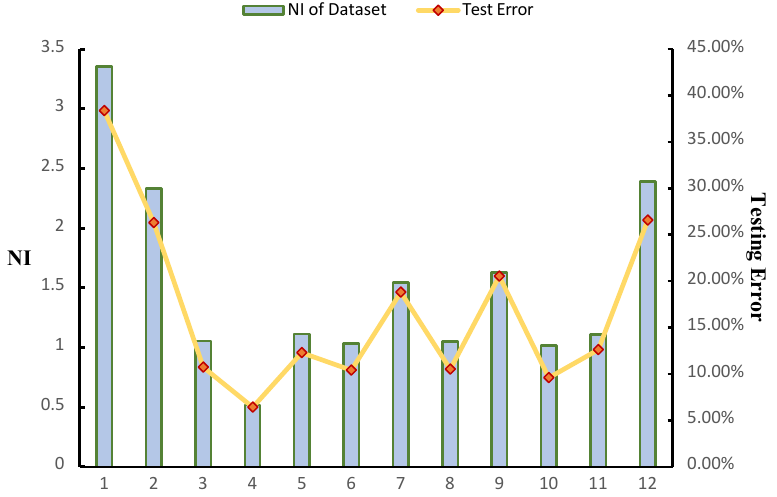}%
    }

    \caption{Comparison of the Index of Distribution Shift and Testing Error.}
    \label{fig3}
\end{figure}

Although the aforementioned APP53 dataset has some scenarios for O.O.D. traffic identification, existing publicly available O.O.D. datasets that can be used for evaluation are still scarce. At the same time, most of the publicly available datasets follow the experimental assumption of independent and identical distributions without considering the impact of distributional variations, which lacks explicit support for the evaluation of O.O.D. traffic identification. Datasets for simulating different O.O.D. traffic scenarios to better support traffic classification studies are still vacant.

In order to more efficiently support and promote the research of O.O.D. encrypted traffic analysis, we design and construct NETD (Dynamic Non-I.I.D. Encrypted Traffic Dataset), an O.O.D. encrypted traffic dataset that supports dynamically adjustable distributions, within the existing publicly available conventional encrypted traffic dataset, ISCX-VPN \cite{Draper-GilLMG16}. We construct NETD by exploiting the intrinsic variability of target concepts and background contexts in the traffic data to achieve distributional shifts, \emph{i.e.}, to simulate the distributional changes brought about by the development of objective factors such as time. By treating any network behaviour as a contextual principal component and other network behaviours as secondary components in the traffic data, the degree of flexibility in controlling the distributional shift can be achieved by adjusting the proportional deviation of the principal and secondary components in the target traffic task.

Given a feature extractor $G$ as well as a category $C$, we introduce the Non-I.I.D. Index (NI) \cite{HeSC21} used to evaluate the impact of distributional shift on the dataset as follows:
\begin{equation}\label{eq1}
	\mathrm{NI}(C) = \left \| \frac{\overline{G(X_{Train}^C)} - \overline{G(X_{Test}^C)}}{\sigma (G(X^C))} \right \| _2
\end{equation}
where $X$ denotes the full data set and $X_C = X_{Train}^C \cup X_{Test}^C$. The $\overline{(\cdot)}$ represents the first-order moments, meaning that the expected or mean value is calculated for the representation of the training set or the test set, which is able to systematically portray the probability distribution of the data set. The $\sigma (\cdot)$ represents the standard deviation, which is used to normalize the dimension of the representation. The $\parallel \cdot \parallel_2$ represents the L2 norm, which is able to measure the degree of difference in the distribution between the train and test sets.

To verify the prevalence of Non-I.I.D. in existing datasets, we use the ET-BERT model as a feature extractor and test it on the widely used I.I.D. dataset ISCX-VPN. Figure \ref{fig3} illustrates the impact of distributional shifts on traffic identification, where traffic classes with more severe distributional shifts produce correspondingly larger identification errors in testing, and the strict independent identity distribution is difficult to satisfy, \emph{i.e.}, few classes are able to achieve a NI value of zero.

\subsubsection{Downstream Tasks}
In accordance with Table \ref{tab:datasets}, we present six encrypted traffic classification tasks:

\textbf{Task 1:} \textbf{E}ncrypted \textbf{A}pplication \textbf{C}lassification with the Same \textbf{T}ime Distribution (EAC-T) aims to set up and classify the identification of application traffic that is under the same time.

\textbf{Task 2:} \textbf{E}ncrypted \textbf{A}pplication \textbf{C}lassification for the Same Application \textbf{V}ersion (EAC-V) aims to classify application traffic collected from the same version of the application.

\textbf{Task 3:} \textbf{E}ncrypted \textbf{A}pplication \textbf{C}lassification with \textbf{T}ime Shift (EAC $\Rightarrow$ T) aims to classify application traffic based on time span (one month interval). Task 1 will be fine-tuned under this task and tested in the form of zero-shot on traffic collected at different times.

\textbf{Task 4:} \textbf{E}ncrypted \textbf{A}pplication \textbf{C}lassification with \textbf{V}ersion Shift (EAC $\Rightarrow$ V) aims to categorise application traffic based on version span (version update). Task 2 will be fine-tuned under this task and tested in the form of zero-shot on traffic data collected from another application version.

\textbf{Task 5:} \textbf{M}alicious \textbf{S}ervice \textbf{C}lassification with \textbf{T}ype Shift (MSC $\Rightarrow$ T) aims to classify the botnet traffic with type shifts. In this task, benign traffic is one class, while botnet traffic consists of 16 types, of which the training set contains only a portion of seven of them. In addition to the default binary classification, we add a multi-classification scenario.

\textbf{Task 6:} \textbf{E}ncrypted \textbf{T}raffic \textbf{C}lassification with Distribution \textbf{F}lexible (ETC $\Rightarrow$ F) aims to classify the encrypted traffic under different distributional variations. In this task, twelve web services are grouped into four different distributional datasets through different shifts in the context.

\begin{table}[t]
  \centering
\caption{Category Distribution of the Datasets.}
\label{tab:dataset-category}
\begin{tabularx}{\linewidth}{>{\centering\arraybackslash}m{2.25cm}|>{\centering\arraybackslash}m{0.8cm}|>{\raggedright\arraybackslash}X}
    \toprule
    Dataset & \#Label & Category \\
    \midrule
    \begin{tabular}[c]{@{}c@{}}APP53-TIME\\APP53$\Rightarrow$TIME\end{tabular} & {28} & {air.ITVMobilePlayer},
    {bbc.mobile.weather},
    {cn.cntv},
    {com.amazon.kindle},
    {com.booking},
    {com.cnn.mobile.android.phone},
    {com.dropbox.android},
    {com.flipkart.android},
    {com.google.android.gm},
    {com.groupon},
    {com.gumtree.android},
    {com.iconology.comics},
    {com.imdb.mobile},
    {com.imo.android.imoim},
    {com.joelapenna.foursquared},
    {com.lenovo.anyshare.gps},
    {com.pinterest},
    {com.reddit.frontpage},
    {com.shazam.android},
    {com.shpock.android},
    {com.skype.raider},
    {com.soundcloud.android},
    {com.ted.android},
    {com.tripadvisor.tripadvisor},
    {info.androidz.horoscope},
    {kik.android},
    {sg.bigo.live},
    {tv.twitch.android.app} \\
    \midrule
    \begin{tabular}[c]{@{}c@{}}APP53-VERSION\\APP53$\Rightarrow$VERSION\end{tabular} & {25} &
    {com.amazon.mShop.android.shopping},
    {com.badoo.mobile},
    {com.contextlogic.wish},
    {com.facebook.katana},
    {com.google.android.youtube},
    {com.guardian},
    {com.ideashower.readitlater.pro},
    {com.instagram.android},
    {com.jingdong.app.mall},
    {com.kakao.talk},
    {com.nytimes.android},
    {com.particlenews.newsbreak},
    {com.sina.weibo},
    {com.snapchat.android},
    {com.spotify.music},
    {com.ss.android.ugc.aweme},
    {com.taobao.taobao},
    {com.tencent.qqlivei18n},
    {com.twitter.android},
    {com.vidio.android},
    {com.vkontakte.android},
    {com.xunmeng.pinduoduo},
    {flipboard.app},
    {ru.ideast.championat},
    {tv.danmaku.bili} \\
    \midrule
    \begin{tabular}[c]{@{}c@{}}{ISCX-Botnet}\end{tabular} & \makecell[c]{2(8)} & Benigh and Malicious (Neris, Rbot, Virut, NSIS, SMTP Spam, Zeus, Zeus Control) \\
    \bottomrule
  \end{tabularx}
\end{table}

\textbf{Notes}. In our analysis of the APP53, we identified inconsistencies between the application categories linked to the shift factors and the descriptions in the original work, along with difficulties in obtaining fully accurate labels. In APP53, version changes involve 25 application categories, not 22 as originally described, and labeling inconsistencies contributed to lower test accuracy. Additionally, in ISCX-Botnet (Multi-Class), only 5 classes could be mapped to known labels. Thus, we supplemented the remaining two categories based on the original combined datasets. As a result, we conducted the experiments in this paper based on the actual dataset acquisition to ensure accurate evaluation results. The category distribution can be seen in Table \ref{tab:dataset-category}.

Echoing these tasks, we implement experiments to validate the effectiveness of our framework in a variety of settings and address key research questions:

\begin{itemize}
  \item \textbf{RQ1:} How does the proposed ETooL framework perform in
both supervised and zero-shot settings for traffic classification? (Section \ref{sec:exp1})
  \item \textbf{RQ2:} What is the contribution of various key components in the proposed ETooL framework to its overall performance? (Section \ref{sec:exp2})
  \item \textbf{RQ3:} What is the generalization ability of our model in handling dynamic distribution shift? (Section \ref{sec:exp3})
  \item \textbf{RQ4:} How efficient is the ETooL framework? (Section \ref{sec:exp4})
  \item \textbf{RQ5:} What extent does hyper-parameter selection affect the results? (Section \ref{sec:exp5})
\end{itemize}

\subsubsection{Comparison Methods}
The state-of-the-art (SOTA) methods used by application fingerprinting are summarized as comparison approaches, including (1) statistical feature methods: AppScanner \cite{TaylorSCM18} and CUMUL \cite{PanchenkoLPEZHW16}; (2) deep learning methods: Deep Fingerprinting (DF) \cite{Sirinam2018}, FS-Net \cite{LiuHXCL19} and GraphDApp \cite{ShenZZXD21}; (3) pre-training methods: PERT \cite{HeYC20} and ET-BERT \cite{LinXGLSY22}. These methods are selected because they represent a subset of different technical approaches that support flow-level traffic detection. They are representative, widely adopted, and commonly used as baselines in comparative studies.

\subsubsection{Evaluation Metrics}
For each experiment, we evaluate the methods by four typical metrics, including Accuracy ($AC=(TP+TN)/(TP+TN+FP+FN)$), Precision ($PR=TP/(TP+FP)$), Recall ($RC=TP/(TP+FN)$), and F1-Score($F1=(2*PR*RC)/(PR+RC)$). Macro Average \cite{LiuWWLM17} is used to avoid biased results due to imbalance between multiple categories of data by calculating the mean value of PR, RC and F1 of each category.
\begin{equation}
\begin{aligned}
\text{Macro-Precision} &= \frac{1}{N}\sum_{i=1}^{N} \text{Precision}_{i},\\[4pt]
\text{Macro-Recall}    &= \frac{1}{N}\sum_{i=1}^{N} \text{Recall}_{i},\\[4pt]
\text{Macro-F1}        &= \frac{1}{N}\sum_{i=1}^{N} \text{F1}_{i}.
\end{aligned}
\end{equation}

\subsubsection{Implementation}
We employ Vicuna-7B-v1.5 as the base model for our approach. Unlike traditional Transformers, this model is optimized by several key modifications: replacing LayerNorm with RMSNorm, Multi-Head Attention with Grouped-Query Attention, Positional Encoding with Rotary Position Embedding, and ReLU with SwiGLU as the activation function. The architecture consists of 32 decoder layers, each with 32 self-attention heads, and the dimensions of the q, k, and v vectors in the attention module are set to 128. ETooL is set with the learning rate of $2 \times e^{-3}$, a warmup ratio of $3 \times e^{-2}$, the training epoch of 3, the batch size of 2 and the maximum input length of the LLM to 2,048. We set the BURST time threshold to a value of 1s. All the experiments are implemented with Pytorch 2.1.0, conducted with NVIDIA Tesla A800 GPUs with 80 GB.

\subsection{Overall Performance Comparison (RQ1)}
\label{sec:exp1}
We conduct experiments on the traffic classification tasks, evaluating
both supervised and zero-shot scenarios. The overall performance is presented in Table \ref{tab:exp-all}. Our ETooL consistently outperforms various state-of-the-art baselines in both supervised and zero-shot scenarios.

\subsubsection{General Encrypted Traffic Classification in I.I.D.}
We first discuss the performance between our proposed model and the existing methods in classifying apps in ideal experimental setting. The experiments in this section also play the role of baseline to indict the effect of ambiguous traffic and concepts drift in the following sections.

Our proposed ETooL achieves an average performance improvement of 5.41\%, 7.49\%, and 19.87\% on F1 compared to different representative traffic classification baselines (ET-BERT, FS-Net, and AppScanner) in I.I.D. scenarios.

In the EAC-TIME task, ETooL achieves performance improvements of 6.62\%, 7.63\%, and 20.92\% in F1 score compared to ET-BERT, FS-Net, and AppScanner, respectively. The APP53 dataset's homogeneous flow interference hampers traditional traffic feature construction methods (such as CUMUL and AppScanner), indicating the diminishing effectiveness of expert-driven feature extraction. DF and FS-Net, which employ deep learning for feature extraction, can recognize encrypted traffic with I.I.D. flows. However, their performance varies due to differences in network architecture and feature selection, and both are constrained by the limited size of labelled data. Although ET-BERT demonstrates stronger recognition performance, confirming the advantage of pre-training over traditional methods, its effectiveness is hindered by the limitations of flow-level input. In contrast, ETooL does not rely on extensive flow pre-training; instead, it leverages the comprehension capabilities of large language models through instruction tuning, exploiting multi-flow correlations and contextual relationships from limited labelled data, thus enabling effective performance in challenging I.I.D. scenarios.

\begin{table*}[!htbp]
  \setlength\tabcolsep{5.5pt}
  \centering
  \caption{Comparison Results on APP53 in I.I.D. and O.O.D. Setting.}
  \label{tab:exp-all}
  \begin{tabular}{c|cccc|cccc|cccc|cccc}
    \toprule
    \textbf{Dataset} & \multicolumn{4}{c|}{\textbf{APP53-TIME}} & \multicolumn{4}{c|}{\textbf{APP53-VERSION}} & \multicolumn{4}{c|}{\textbf{APP53 $\Rightarrow$ TIME}} & \multicolumn{4}{c}{\textbf{APP53 $\Rightarrow$ VERSION}}\\
    \midrule
    \textbf{Method} & AC\%& PR\%& RC\%& F1\%& AC\%& PR\%& RC\%& F1\%& AC\%& PR\%& RC\%& F1\%& AC\%& PR\%& RC\%& F1\%\\
    \midrule
    AppScanner 
    & 72.65 & 72.90 & 72.65 & 72.27 
    & 73.08 & 73.53 & 73.53 & 73.29
    & 43.87 & 46.02 & 43.87 & 44.03 
    & 35.39 & 35.85 & 35.69 & 35.00\\
    CUMUL 
    & 60.13 & 59.93 & 60.38 & 59.74  
    & 61.00 & 60.18 & 61.04 & 60.22
    & 37.78 & 37.91 & 37.78 & 37.46  
    & 36.83 & 36.33 & 37.17 & 36.14\\
    \midrule
    DF
    & 69.30 & 69.44 & 69.21 & 69.10 
    & 73.44 & 74.49 & 73.63 & 73.75
    & 34.56 & 36.71 & 34.56 & 34.56 
    & 34.58 & 34.51 & 34.58 & 33.78\\
    FS-Net
    & 85.75 & 85.91 & 85.46 & 85.56 
    & 84.50 & 84.82 & 84.91 & 84.77
    & 45.24 & 45.11 & 44.87 & 43.96 
    & 43.14 & 46.42 & 43.33 & 43.48\\
    GraphDApp
    & 86.53 & 84.18 & 85.66 & 85.96 
    & 80.32 & 81.21 & 80.93 & 80.47
    & 41.41 & 42.16 & 41.98 & 41.79 
    & 34.41 & 34.32 & 34.26 & 34.24\\
    \midrule
    PERT
    & 78.70 & 78.70 & 79.29 & 78.71 
    & 76.40 & 76.40 & 77.52 & 76.77
    & 45.69 & 45.69 & 49.20 & 46.18 
    & 45.96 & 45.96 & 47.54 & 45.72\\
    ET-BERT 
    & 86.54 & 86.54 & 86.72 & 86.57
    & {87.82} & 87.72 & {87.94} & {87.92}
    & 57.04 & 57.04 & 59.37 & 56.71
    & {57.08} & 57.08 & {60.07} & {56.98}\\
    \midrule
    \textbf{ETooL}
    & \textbf{93.14} & \textbf{94.21} & \textbf{93.77} & \textbf{93.19}  
    & \textbf{93.39} & \textbf{92.66} & \textbf{91.83} & \textbf{92.11}
    & \textbf{74.38} & \textbf{76.24} & \textbf{74.12} & \textbf{74.88}
    & \textbf{73.19} & \textbf{74.24} & \textbf{72.18} & \textbf{72.13}\\
    \bottomrule
  \end{tabular}
\end{table*}

\begin{table}[!htbp]
  \setlength\tabcolsep{3.5pt}
  \centering
  \caption{Comparison Results on ISCX-Botnet in O.O.D. Setting.}
  \label{tab:exp-all-2}
  \begin{tabular}{c|cccc|cccc}
    \toprule
    \textbf{Dataset} & \multicolumn{4}{c|}{\textbf{MSC $\Rightarrow$ T (Binary)}} & \multicolumn{4}{c}{\textbf{MSC $\Rightarrow$ T (Multi-Class)}}\\
    \midrule
    \textbf{Method} & AC\%& PR\%& RC\%& F1\%& AC\%& PR\%& RC\%& F1\%\\
    \midrule
    AppScanner 
    & 76.94 & 76.97 & 76.95 & 76.92 
    & 60.99 & 60.12 & 60.99 & 60.16\\
    CUMUL 
    & 65.95 & 66.77 & 65.88 & 65.42  
    & 29.41 & 56.96 & 29.41 & 28.37\\
    \midrule
    DF
    & 78.25 & 79.90 & 78.28 & 77.94 
    & 26.70 & 44.30 & 42.10 & 42.30\\
    FS-Net
    & 77.42 & 78.81 & 77.36 & 77.27 
    & 70.33 & 71.39 & 71.34 & 69.87\\
    GraphDApp
    & 77.92 & 78.13 & 77.43 & 77.81 
    & 71.36 & 71.29 & 71.37 & 71.31\\
    \midrule
    PERT
    & 85.23 & 88.77 & 85.11 & 85.19 
    & 65.38 & 62.70 & 65.36 & 61.48\\
    ET-BERT 
    & 85.74 & 85.18 & 88.67 & 85.87
    & {65.51} & 72.58 & {65.51} & {67.91}\\
    \midrule
    \textbf{ETooL}
    & \textbf{95.36} & \textbf{95.72} & \textbf{94.87} & \textbf{95.03}  
    & \textbf{84.19} & \textbf{83.02} & \textbf{82.73} & \textbf{81.95}\\
    \bottomrule
  \end{tabular}
\end{table}

In the EAC-VERSION task, ETooL significantly outperforms the three leading methods, with F1 score improvements of 4.19\%, 7.34\%, and 18.82\%, respectively. The increased difficulty of identifying encrypted traffic stems from the larger number of data categories for recording application versions, as well as the interference of similar flows. Despite these challenges, ETooL shows superior robustness, affirming its powerful capacity to comprehend and generalize traffic features effectively.

\subsubsection{Non-I.I.D. Encrypted Traffic Classification}
In this subsection, we explore the generalization ability of our model by incorporating more instruction data to fine-tune the ETooL for effectively handling various types of tasks. In contrast to supervised I.I.D. experiments, this subsection will discuss the out-of-distribution generalisation capabilities of our proposed method under distributional variations.

According to Table \ref{tab:exp-all}, the baseline method shows a significant degradation in performance in the face of changes in flow distribution. Nevertheless, our proposed ETooL achieves the lowest performance degradation and an average performance improvement of 16.66\%, 29.79\%, and 33.42\% on F1 compared to different representative baselines in the APP53 $\Rightarrow$ TIME and APP53 $\Rightarrow$ VERSION tasks, respectively.

In the EAC $\Rightarrow$ TIME task, the traffic identification accuracy of existing representative encrypted traffic classification methods decreases significantly when no out-of-distribution handling strategies are applied. This decline is particularly evident when the methods are tested on new distributions. For instance, AppScanner achieves an F1 score of 72.27\% on I.I.D. traffic, yet its performance drops to 44.03\% when evaluated on test data with time shifts. Similarly, FS-Net and ET-BERT exhibit declines in F1 scores, dropping to 43.96\% and 56.71\%, respectively. These results illustrate that shifts in traffic distribution, caused by changes in time intervals, have a substantial impact on the performance of existing methods.

However, the ETooL framework demonstrates superior robustness, experiencing minimal degradation and exhibiting enhanced O.O.D. traffic identification capabilities compared to the other methods. Unlike these methods, which struggle to adapt to O.O.D. traffic caused by time variation, ETooL is capable of maintaining effective performance. This demonstrates that, despite shifts in the distribution of traffic over time, certain invariant properties persist in O.O.D. traffic. These properties, rooted in the associations within the traffic transport topology and the contextual relationships of traffic flows, can be effectively captured and integrated into ETooL’s inference mechanism, allowing it to generalize and perform well even under time-varying conditions.

In the EAC $\Rightarrow$ VERSION task, we observe that the impact from version updates is relatively stronger than temporal changes, seeing that this is a more challenging task for identifying out-of-distribution encrypted traffic. Unlike time spanning, the cross-version task is to collect encrypted traffic from two different versions of a mobile application at the same time, when differences in application network interfaces or service design logic bring about differences in traffic distribution. AppScanner's traffic identification result drops to 35.00\%, while FS-Net and ET-BERT both drop to 43.48\% and 56.98\% respectively. However, the ETooL model still maintains minimal performance degradation and provides a significant improvement over existing methods.

Furthermore, we validate the performance of ETooL in the detection of malicious traffic. When distinguishing between benign and malicious traffic, a key source of distributional bias arises from the variation among different botnet types present in the malicious traffic. In addition to the variation factors encountered in the two previously mentioned O.O.D. tasks, the MSC $\Rightarrow$ T (Binary) task introduces another layer of complexity by including malicious traffic from previously unknown botnet types. This necessitates a higher degree of model generalization, as the model must effectively generalize beyond the known botnet types to accurately detect and classify these unseen forms of malicious traffic. Based on the characteristics of the dataset, we also incorporated benign traffic alongside seven types of botnet traffic to perform a multi-class classification task for malicious traffic detection.

As shown in Table \ref{tab:exp-all-2}, AppScanner and FS-Net, both of which rely on packet size as a feature, exhibit superior detection rates compared to their counterparts (excluding unknown botnet types). This suggests that packet size is a valuable feature for capturing commonalities across distributional shifts. Additionally, the use of datagrams as feature carriers in DF, compared to pre-training methods such as PERT and ET-BERT, further supports the performance gains attributable to the pre-training architecture. Notably, ETooL surpasses the optimal baseline by 9.16\% and 12.08\% across two malicious traffic detection tasks. On the one hand, ETooL leverages TRG to fuse packet size and datagram features, alongside incorporating concurrency and timing relationships of flows. This allows the model to capture richer underlying interactions within the traffic data. Second, ETooL’s large-scale architecture enhances its inference capabilities through pre-training, while also improving its understanding of graph structures. These factors collectively enable ETooL to generalize more effectively to unseen traffic distributions.

\begin{table*}[!htbp]
  \centering
  \caption{Ablation Study of Key Components in ETooL.}
  \label{tab:exp-ab}
  \begin{tabular}{c|c|cc|cc|cc|cc|cc}
    \toprule
    \multicolumn{2}{c|}{\textbf{Dataset}} & \multicolumn{2}{c|}{\textbf{APP53-T}} & \multicolumn{2}{c|}{\textbf{APP53-V}}& \multicolumn{2}{c|}{\textbf{APP53 $\Rightarrow$ T}} & \multicolumn{2}{c|}{\textbf{APP53 $\Rightarrow$ V}} &
    \multicolumn{2}{c}{\textbf{Botnet $\Rightarrow$ T}} \\
    \midrule
    \multicolumn{2}{c|}{\textbf{Method}} & AC\%& F1\%& AC\%& F1\%& AC\%& F1\%& AC\%& F1\%& AC\%& F1\%\\
    \midrule
    \multicolumn{2}{c|}{\textbf{ETooL (Full Model)}}
    & \textbf{93.14} & \textbf{93.19} & \textbf{93.39} & \textbf{92.11} & \textbf{74.38} & \textbf{74.88} & \textbf{73.19} & \textbf{72.13} & \textbf{95.36} & \textbf{95.03} \\
    \midrule
    1 & \emph{w/o} Raw Datagram
    & 90.93 & 89.62 & 89.33 & 88.64 & 68.76 & 68.18 & 67.26 & 67.93 & 93.26 & 92.83 \\
    2 & \emph{w/o} Packet Length 
    & 91.73 & 90.62 & 91.31 & 90.99 & 69.43 & 69.39 & 70.55 & 69.72 & 94.75 & 94.39 \\
    \midrule
    3 & \emph{w/o} Graph Structural Tuning 
    & 74.25 & 73.92 & 76.74 & 76.81 & 45.87 & 45.79 & 43.56 & 43.11 & 77.83 & 77.02 \\
    4 & \emph{w/o} Large Language Model 
    & 87.26 & 86.95 & 87.01 & 85.72 & 55.61 & 54.79 & 52.20 & 51.96 & 83.47 & 83.69 \\
    \bottomrule
  \end{tabular}
\end{table*}

\subsection{Ablation Study (RQ2)}
\label{sec:exp2}
We conduct an ablation study to investigate the individual contributions of different sub-modules of our proposed framework, and the results are reported in Table \ref{tab:exp-ab}.

We sequentially eliminate raw datagram, packet length, graph structural tuning, and LLM and show the ablation results to verify the contribution of each component on different tasks.

(1) In Models "1-2", we evaluate the impact of different granularities of traffic input information on the model's effectiveness. Model "1", which excludes datagram sequences, shows an average decrease of 4.02\% in F1 score compared to ETooL. Similarly, removing packet length sequences results in an average decrease of 2.44\%. These findings suggest that while the representation of arbitrary traffic information provides some representational gain, datagrams offer more significant improvements in model performance.

(2) In Model "3", we assess the impact of the traffic graph structure and the graph instruction fine-tuning task. In this model, flow-level instruction fine-tuning is performed directly using the large language model, without incorporating traffic relation graphs. Model "3" exhibits significant performance degradation across all scenarios, with an average reduction of 22.13\% in F1 score compared to the full model. These results suggest that both the traffic correlation structure and the graph instruction tuning paradigm are critical for enabling ETooL to learn traffic context more effectively. Furthermore, this interaction-based approach enhances the model’s ability to capture representational similarities under distributional shifts, thereby improving its performance in O.O.D. traffic detection.

(3) Model "4" was designed to perform supervised training on flow graphs using an uninitialized Graph Transformer model, removing the instruction tuning paradigm, and then testing its out-of-distribution capability. Compared to ETooL, Model "4" demonstrates an average F1 score decrease of 12.84\% across all datasets. This indicates that the understanding and reasoning capabilities provided by the large language model play a critical role in mitigating misclassification after distributional shifts, offering substantial support for the task of O.O.D. traffic identification.

\subsection{Generalization Ability Investigation (RQ3)}
\label{sec:exp3}

To further measure the performance difference between the ETooL model and the comparative approach models, we analyze the encrypted traffic classification capability under the ISCX-VPN with I.I.D. setups and the dynamic distribution-variation traffic dataset NETD.

As described in Section \ref{sec:dataset}, we further illustrate the detailed construction process. NETD is primarily designed by controlling two key factors: proportional bias and compositional bias. The specific settings are as follows:

\textbf{Basic Dataset Composition}. The ISCX-VPN dataset consists of 6 types of services under both VPN and Non-VPN categories, encompassing a total of 17 applications, namely Chat (ICQ, AIM, Skype, Facebook and Hangouts), Email (SMPTS, POP3S and IMAPS), File Transfer (Skype, FTPS and SFTP), P2P (uTorrent and Transmission), Streaming (Vimeo and Youtube), VoIP (Facebook, Skype and Hangouts). With the detection objective of service traffic identification in mixed traffic scenario, the components of the constituent services are applied.

\begin{figure}[t]
    \centering
    \subfloat[NETD-1]{%
        \includegraphics[width=0.5\columnwidth]{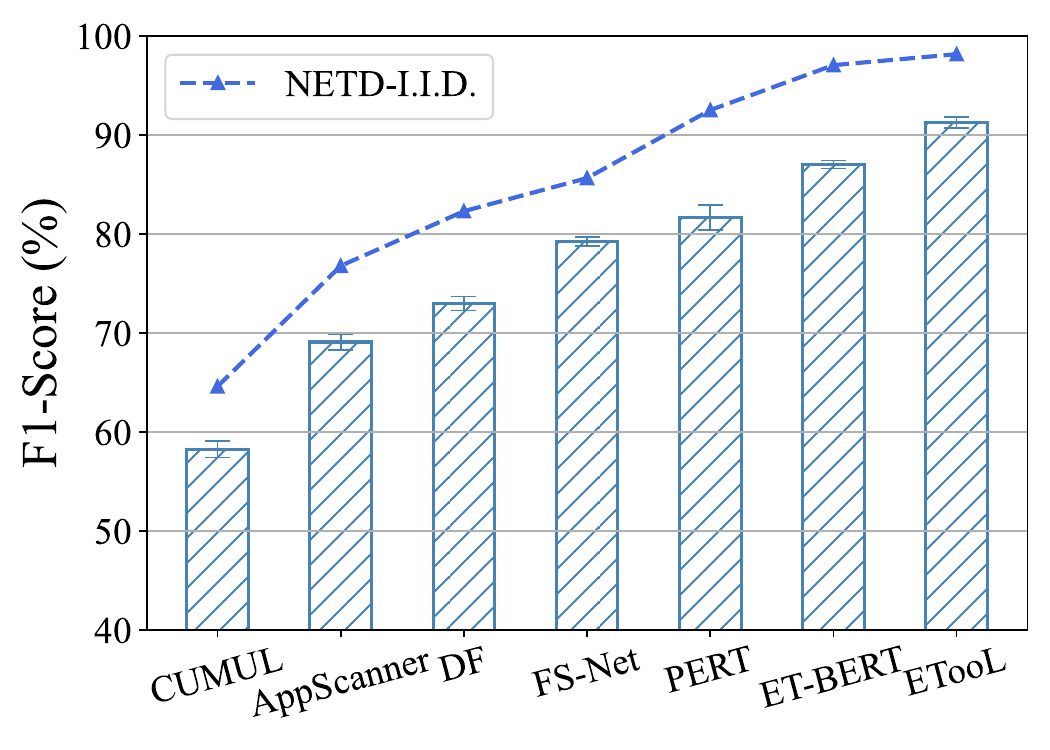}%
    }
    \hfill
    \subfloat[NETD-2]{%
        \includegraphics[width=0.5\columnwidth]{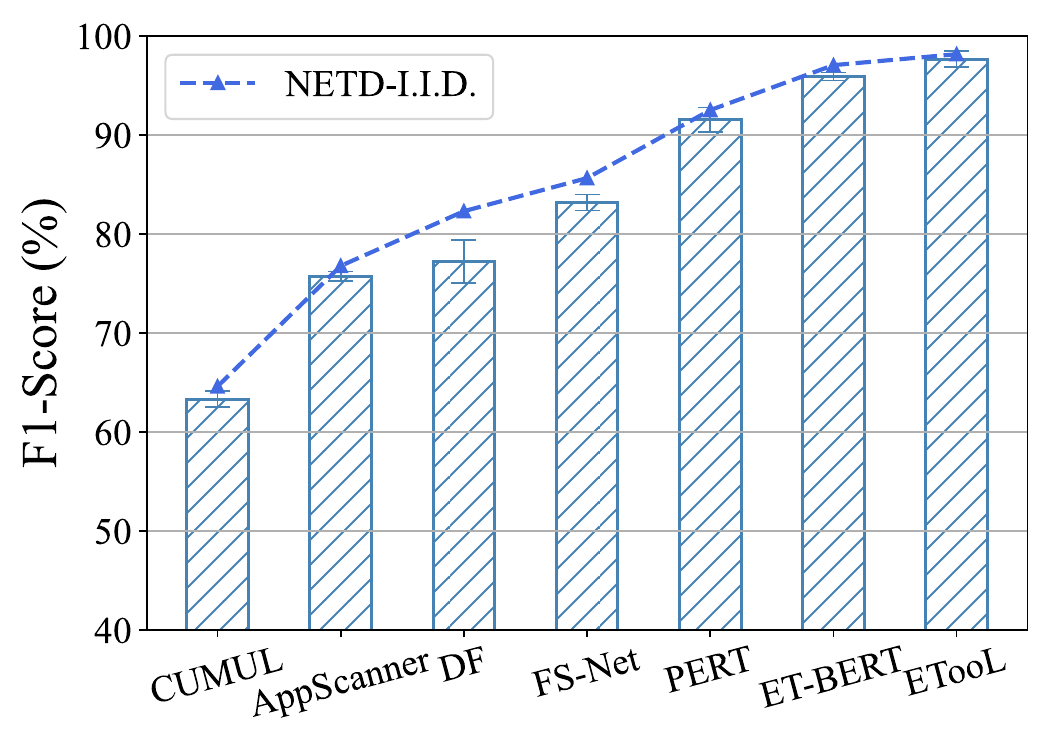}%
    }

    \subfloat[NETD-3]{%
        \includegraphics[width=0.5\columnwidth]{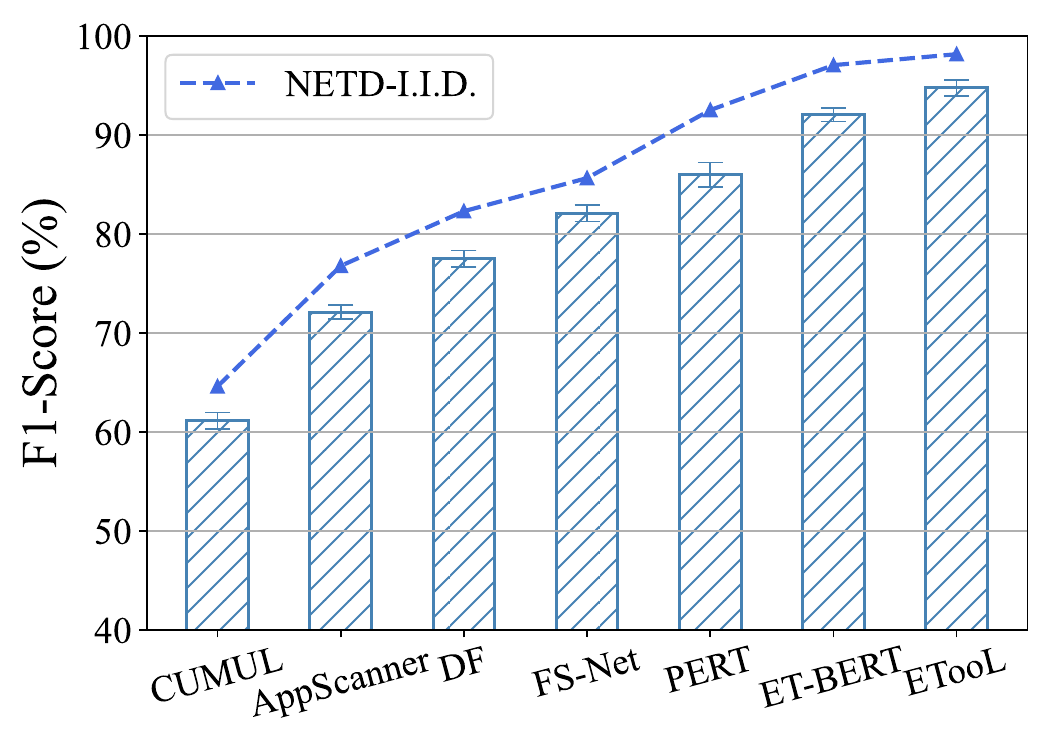}%
    }
    \hfill
    \subfloat[NETD-4]{%
        \includegraphics[width=0.5\columnwidth]{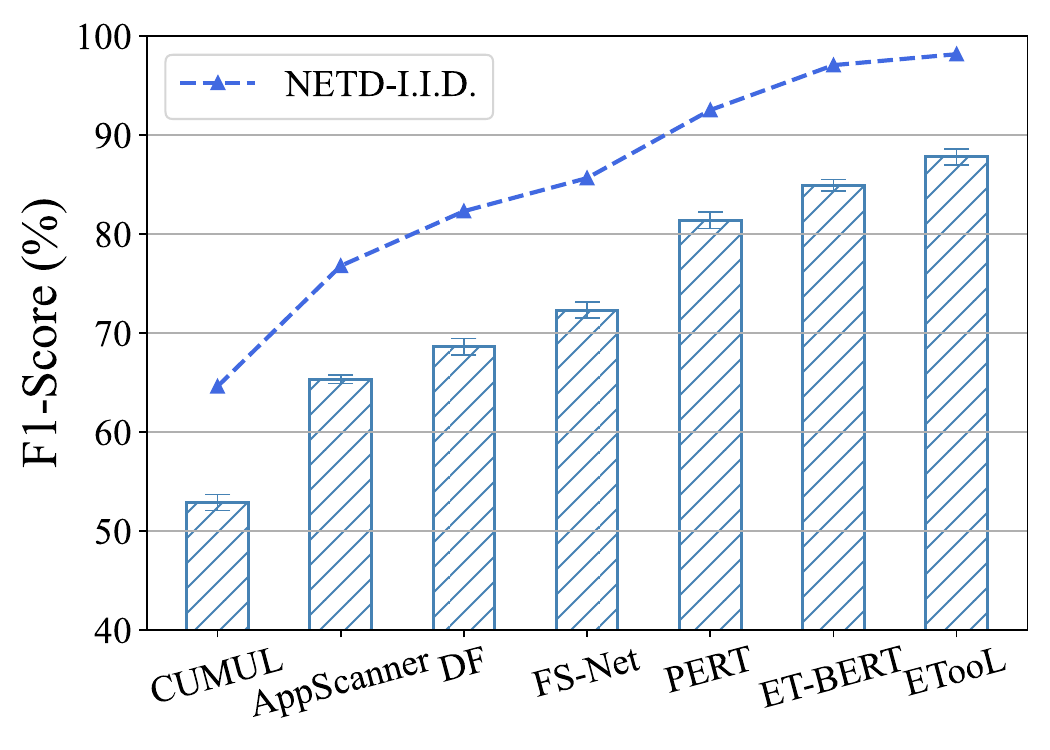}%
    }

    \caption{Comparison Results on Dynamic Non-I.I.D. Encrypted Traffic Dataset.}
    \label{fig4}
\end{figure}

\textbf{Proportional Bias Setting}. In this setting, we ensure that the constituent components of each target class are present in both the training and testing data. For each service category, one primary component is randomly selected. The proportional bias between the primary and other components is controlled by specifying a dominant ratio, which determines the relative prevalence of the primary component:
\begin{equation}
\textit{Dominant Ratio} \;=\; \frac{N_{\text{Dominant}}}{N_{\text{Minor}}}
\end{equation}
where $N_{Dominant}$ denotes the number of samples for the dominant component, while $N_{Minor}$ represents the average number of samples for the minor components. By fixing the dominant ratio in either the training or testing data and varying the proportional bias in the other, we simulate different distribution shift scenarios.

\textbf{Compositional Bias Setting}. In contrast to proportional bias, compositional bias simulates the situation where knowledge in the training data fails to cover the complete distribution. By varying the number of constituent components for each service category in the training and test data, we are able to simulate different degrees of information loss and thus achieve distributional bias. For the set of contextual components $C'$, the construction strategy for the training and testing set is as follows:
\begin{equation}
\begin{aligned}
\mathcal{T} = \Bigl\{\,T \subset \mathcal{C'}\ \Bigm|\ 1 \le |T| \le N-1 \Bigr\},\\
|\mathcal{T}| = \sum_{k=1}^{N-1} \binom{N}{k} = 2^{N} - 2
\end{aligned}
\end{equation}
\begin{equation}
\begin{aligned}
\mathcal{S}
  = \Bigl\{\,S \subset \mathcal{C'}\ \Bigm|\ 1 \le |S| \le N \Bigr\},\\
  |\mathcal{S}| = \sum_{m=1}^{N} \binom{N}{m} = 2^{N} - 1
\end{aligned}
\end{equation}
where $\mathcal{T}$ denotes the optional set of training data, $\mathcal{S}$ denotes the optional set of testing data, and $N$ denotes the full number of contextual components of the current category.

\begin{table*}[t]
  \centering
  \caption{Study on the Time and Space Efficiency of the Model Training and Inference.}
  \label{tab:exp4}
  \begin{tabular}{c|c|c|c|c|c|c}
    \toprule
    \multicolumn{2}{c|}{\textbf{Fine-tuning Method}} & \textbf{Training Time} & \textbf{Tuned Parameters}& \textbf{GPU(MB)}  & \textbf{FLOPs} & \textbf{Inference Latency(ms)} \\
    \midrule
    1 & BURST Graph Matching tuning
    & OOM & 6,607,884,288 & OOM & 6,704,861,184 & - \\
    2 &  BURST Graph Matching freeze
    & 18:36:27 & 131,612,672 & 73,511 & 2,915,157,037 & 3,756.62 \\
    \midrule
    3 & Traffic Task tuning 
    & OOM & 6,607,884,288 & OOM & 6,704,861,184 & - \\
    4 &  Traffic Task freeze
    & 1:38:41 & 131,612,672 & 71,196 & 2,793,692,160 & 3,756.62 \\
    \bottomrule
  \end{tabular}
\end{table*}

On the basis of ISCX-VPN, we construct generate four O.O.D. traffic datasets with different distribution shifts: (1) NETD-1: The distribution of the test set and training set of traffic data is changed by using a proportional bias strategy, and the dataset is generated by randomly sampling according to the ratio of major and minor components of 1:3. (2) NETD-2: Similar to NETD-1, but randomly sampling according to a 3:1 ratio of major and minor components in the sample pool. (3) NETD-3: The distribution of the traffic data training set is changed by employing a contextual compositional component bias strategy and tested on the full data set. We randomly capture 80\% of the applications in the target class of services as contextual constituents, while other applications do not appear in the training set. It is also possible to further construct the training data with more severely shifted distributions according to the ratio bias of the major and minor components. (4) NETD-4: Similar to NETD-3, but captures 20\% of the contextual applications of the target service.

According to the results in Fig. \ref{fig4}, the visualization clearly demonstrates that ETooL achieves the best classification performance across various datasets. In the figure, the recognition results for each method are represented by bars, while the folded lines indicate performance under the I.I.D. setting. Notably, ETooL and ET-BERT show comparable performance in the I.I.D. setting. However, under two distinct Non-I.I.D. scenarios, ETooL significantly outperforms the other methods, demonstrating superior robustness and classification accuracy.

\subsection{Model Efficiency Study (RQ4)}
\label{sec:exp4}

The study aims to evaluate the computational efficiency of our model during the training stage. 

As shown in Table \ref{tab:exp4}, our instruction tuning framework follows a two-stage process in which both LLM and graph encoder parameters are frozen and only the flow-graph aligned projection layer is tuned. We perform a comparison between freezing and tuning LLM parameters in a dual-card 80G Nvidia A800 environment, denoted by \textit{freeze} and \textit{tuning}, respectively. The study investigates the time and space efficiency in terms of training time, tuning parameters, single GPU memory occupied (MB), model computing volume and inference latency (milliseconds per response). Under the same experimental conditions, we suffer from GPU Out of Memory errors when fully parametrically tuning a large language model even with one batch size. However, by using a parameter-freezing tuning strategy, the training process can still be executed normally when increasing the training batch size. In addition, the parameters involved in instruction tuning is reduced by more than 50 times with frozen compared to full-parameter, resulting in a significant reduction in model computation and training time. To further investigate the inference efficiency of ETooL, we measured the inference latency on the NETD dataset, using a single NVIDIA A800. While ETooL has not yet met the requirements for real-time detection, it is well-suited for scenarios such as assisted decision-making, where accurate O.O.D. encrypted traffic identification is critical. In particular, when combined with interpretability strategies \cite{HanWFJCWW0SYL24}, LLM's generalization capabilities can be regularized, making it suitable for online deployment.

\begin{figure}[t]
    \centering
    \subfloat[Impact of BURST Time Threshold]{%
        \includegraphics[width=0.5\columnwidth]{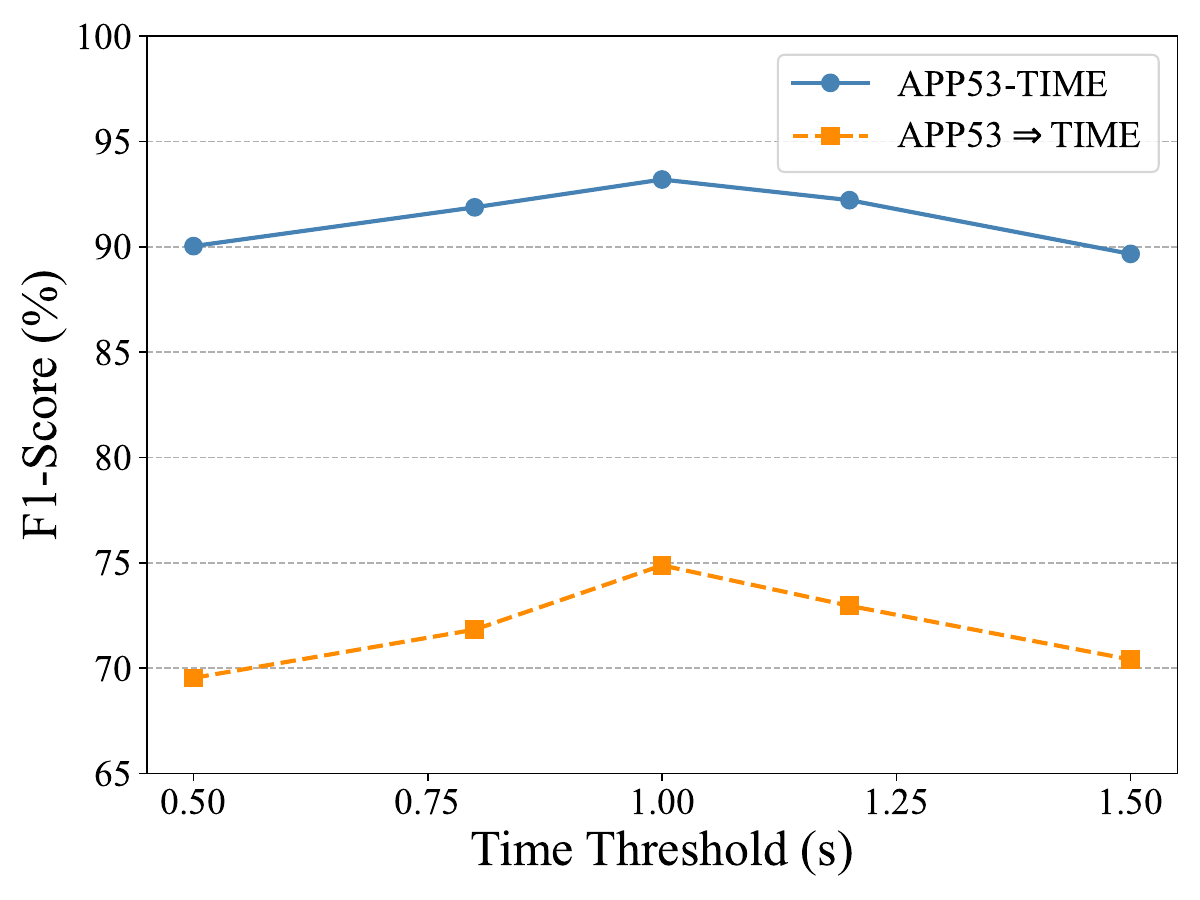}%
    }
    \hfill
    \subfloat[Impact of Learning Rate]{%
        \includegraphics[width=0.5\columnwidth]{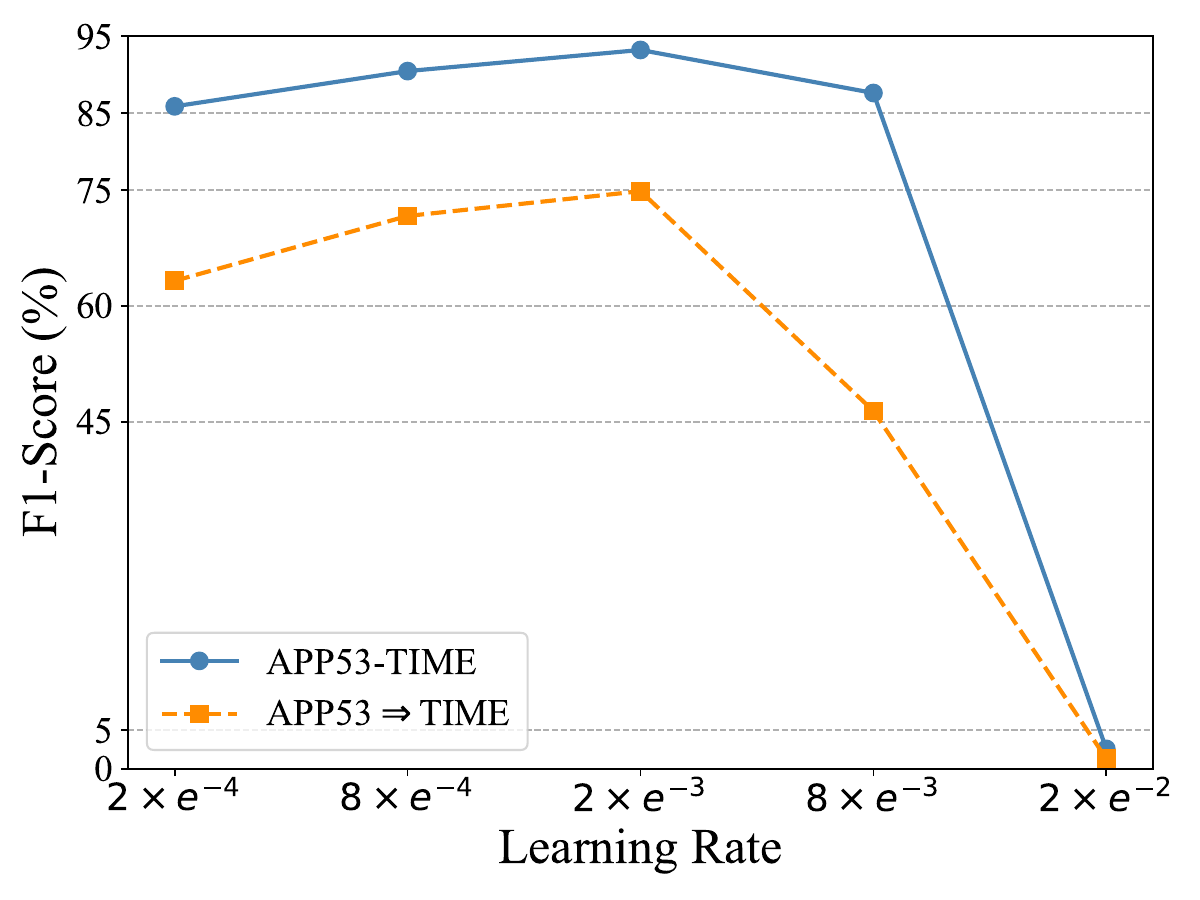}%
    }
    \caption{Comparison Results on Different Hyper-parameters Selection.}
    \label{fig5}
\end{figure}

\subsection{Hyper-parameters Analysis (RQ5)}
\label{sec:exp5}

The analysis aims to evaluate the selection of hyper-parameters during the training stage. According to the results presented in Figure \ref{fig5}, the BURST time threshold and the learning rate affect the performance of the model in the test scenarios of I.I.D. and O.O.D.

If the BURST time threshold is too small, flows serving different functionalities may be aggregated into the same BURST, while an overly large threshold may also lead to the merging of flows with unrelated functions. To evaluate this effect, we experimented with a range of time thresholds and observed the testing performance. Empirically, a threshold of around 1 second yielded the most favorable results.

Moreover, due to the limitation on batch size, setting the learning rate too high can cause the model to oscillate or overshoot near the convergence point, leading to increased gradient variance and potential divergence. On the other hand, setting the learning rate too low slows down convergence, requiring more training steps to compensate. Through experiments across different learning rate ranges, we found that setting the learning rate to $2 \times e^{-3}$ yields the best performance.

\section{Conclusion}
\label{sec:conclusion}
In this paper, we propose ETooL, an effective and distributionally adaptive traffic large language model, aiming to improve the generalisation ability of traffic classification model. The proposed framework injects traffic graph structures based on flow interaction knowledge into the LLM tuning paradigm. We comprehensively evaluate the generalisation ability of ETooL on seven encrypted traffic datasets in I.I.D. and Non-I.I.D. settings, demonstrating the effectiveness of our approach in both supervised and zero-shot scenarios. The experimental results clearly demonstrate that our proposed method exhibits superior out-of-domain generalization capabilities compared to existing encrypted traffic classification approaches. The ETooL framework effectively integrates traffic features and interaction correlation patterns with adaptive instruction tuning via large language models. This enables ETooL to identify out-of-distribution traffic while retaining knowledge of traffic from previous distributions, offering a significant advantage over traditional models that rely on iterative retraining.

\section*{Acknowledgments}
The authors would like to express their grateful appreciation to the associate editor and the anonymous reviewers for their valuable efforts in greatly improving this article.

\vfill

\end{document}